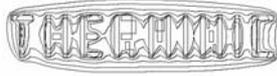

# OPTIMISED CURING OF SILVER INK JET BASED PRINTED TRACES

Z. Radivojevic[1], K. Andersson[1], K. Hashizume[2], M. Heino[1], M. Mantysalo[3], P. Mansikkamaki[3], Y. Matsuba[4], N. Terada[4]

[1] Nokia Research Center, Helsinki, Finland; [2] Nokia Research Center, Tokyo, Japan; [3] Tampere University of Technology, Finland
[4] Tsukuba Research Laboratory Harima Chemicals, Inc, Japan

## ABSTRACT

Manufacturing electronic devices by printing techniques with low temperature sintering of nano-size material particles can revolutionize the electronics industry in coming years. The impact of this change to the industry can be significant enabling low-cost products and flexibility in manufacturing. implementation of a new production technology with new materials requires thorough elementary knowledge creation. It should be noticed that although some of first electronic devices ideally can be manufactured by printing, at the present several modules are in fact manufactured by using hybrid techniques (for instance photolithography, vapor depositions, spraying, etc...). However, in future advances in printing technology may enable the printing of major part of the future electronic modules and interconnections. The main challenge of such opportunity is to provide sufficient quality of interconnecting traces by appropriate materials manipulation and sintering, more precisely appropriate material deposition, process control and sufficient electric conductivity of printed interconnections. In this paper optimal curing conditions of Ag-nano ink deposited by ink-jet printing technique is elaborated. To achieve good product quality, minimal stresses and adequate energy deposition (for sintering of metallic nano particles) have to be understood and optimized. In other words optimal curing technique and energy deposition have to be combined in such a way that stresses are minimized in the whole assembly and manufacturing process. Theoretically minimal stress is obtained if localized energy deposition (only to the nano-ink traces) is used. However, localized heating is challenging requirement for mass applications (slow and low volume process). More adequate solutions take advantage of volumetric curing by optimized thermal management of entire assembly. In such constellation the heat propagation has to be well understood and the whole process optimized.

In this work we investigated the effects of curing conditions on the materials micro structure and electrical conductivity of the cured nano-ink traces. Experimental part involved a set of test vehicles for resistance measurements and variation of energy depositions. A set of nano-ink traces, different in length and thickness were exposed to different curing conditions. Quality of the sintered traces was examined by structural, electrical and numerical methods. By drawing correlations between relevant parameters optimal curing profile for the Ag-nano ink was found. Thermo mechanical modeling and simulations and adequate stress analysis were performed by using ANSYS software.

## INTRODUCTION

Printable electronics presents new promising technology area, which may pave the way to many new low-cost products. Feasible products based on printable electronics might include ultra cheap radio-frequency identification tags, inexpensive and disposable displays/electronic paper, interior interconnections, parts of an electronics assembly (e.g PWB, phone chassis etc..), sensors, memories, and wearable user interfaces. Different materials ranging from metallic inks towards organic materials can be used to produce printed electronic circuits or modules. A key advantage is the ease and speed of fabrication, which simplifies device manufacturing and lowers cost. Ideally, printed electronic circuits can be processed from solution by using similar processes as well-known printing of ink on paper or foils [1]. In addition, as the temperature used in the production process is relatively low, various substrates can be used including cheap, lightweight, and flexible plastic instead of glass [2].

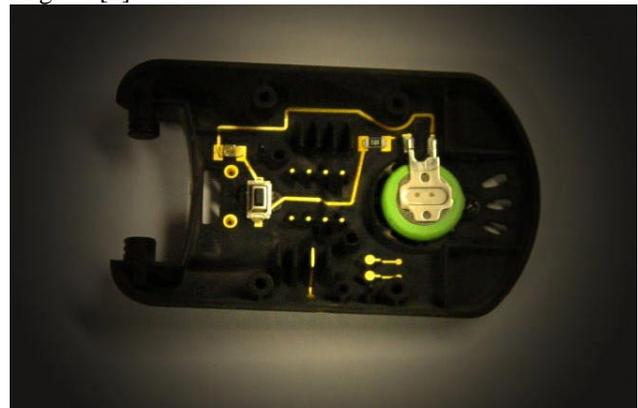

Figure 1. Direct routing lines integrated at mechanical parts of an electronic assembly.

The production of high performance devices by state-of-art technology under high temperatures in clean rooms put a limit on the cost and speed of manufacturing. In addition printed electronics presents a good way towards fast modifications and fast product tailoring to user specific needs (product personalization).

To take full advantage of printing some parts of electronic components with sufficient quality of printed traces sintering of the printed traces has to be well examined and optimal conditions found. This is especially challenging with high density interconnections (below 50 µm). One example of printed routing lines at mechanical body interior of an electronic module (phone) is given in Fig 1.

## EXPERIMENTAL

### Nano Ink

The silver nano ink (60%wt Ag) produced by Harima Chemicals, Inc., by Gas Evaporation Process has been used for the evaluation. This ink is of high quality exhibiting good dispersion and very even size distribution of Ag nano particles with very sharp particle size dispersion with mean value of about 5 nm, as shown in Fig 2.

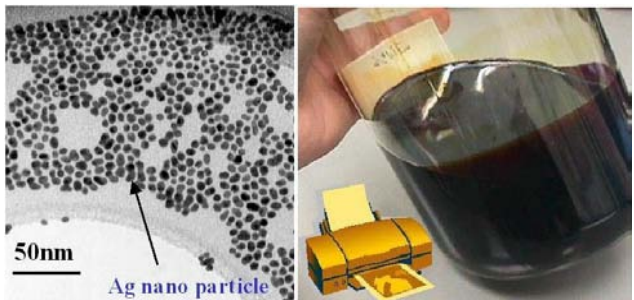

Figure 2. Left: TEM picture of the nano-ink with mean particle size distribution around 5 nm. Right: Ag-nano-ink in liquid form ready for inkjet deposition.

### Deposition

Fine-lines of Ag-nano ink were created on a polymeric substrate by generation of micro droplets using piezoelectric drop-on demand (DOD) printing system. (made by Ricoh Printing Systems, Ltd.). The DOD jetting system is composed of a backpressure controller, a purging system and a piezoelectric jetting system. About 0.9 kP vacuum is maintained in the reservoir to prevent leaking of the nano-ink from the nozzle of the capillary tube due to the small viscosity and low surface tension of the nano-ink. A vacuum controller and a magnetic valve were connected between the vacuum pump and the reservoir to minimize the loss of ink solvent due to continuous evaporation. Nitrogen gas with controlled pressure was used to purge the clogging. Highly miniaturized printhead with a nozzle diameter of 30 µm was nozzle diameter is used to generate micro-droplets. After generating stable droplets of 30 microns at 30 kHz, continuous lines were made on a substrate by moving precision translation stage at 20 mm/ s (see Fig 3). A set of test vehicles with different conducting traces (length, width, thickness) was made by multi-pass approach providing precise adjustment of the width and thickness. The gap between the jetting head tip and the substrate was maintained at 0.4-0.6 mm.

After the deposition of the printed lines the samples were placed in the IrF chamber, nitrogen flow was established and inner pressure was kept at about 1.1. atm.

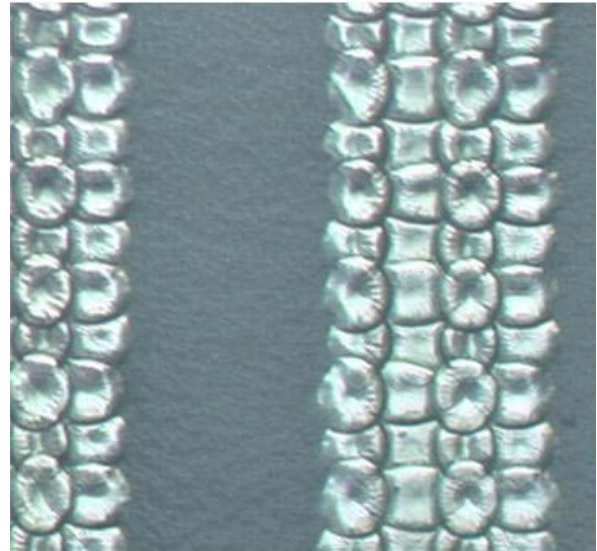

Figure 3. Typical deposited pattern of Ag-nano-ink on substrate.

### Curing

Modern electronics uses a bill of materials with different physical properties. Such set of materials apposes additional conditions for the overall sintering procedure. At one side the sintering has to succeed in good quality of the printed traces (e.g. electric conductivity, physical reliability, material compatibility). On the other side the sintering conditions has to be kept within certain limits to minimize overall plastic deformations and stresses which can deteriorate the system reliability. In other words the long-term reliability of the final product must not be questioned by curing conditions.

In our work the curing has been performed by using electromagnetic IrF View Heating Oven [3] as shown in Fig 4. Spectral characteristic of the curing oven is given in Fig 4, also. Notice that the peaks of the electromagnetic emission have been adjusted to the ink solvent liquid (300 nm) and the Ag- nano particles (430 nm).

Different curing profiles have been performed. The profiles were arranged as the heating treatment with variation in time length ($t$) and curing temperature ($T$). The "easy" profile (profile 1) was set as T=150$^\circ$C for 60 min followed by temperature elevated at 240$^\circ$C for 30 min. "Medium" profile (profile 2) comprised constant temperature of 240°C for 60 min. Profile 3 included pre heating at 150°C for 60 min followed by 240°C for 60 min. Curing at 300°C for 60 min represented the "hardest" profile (profile 4).

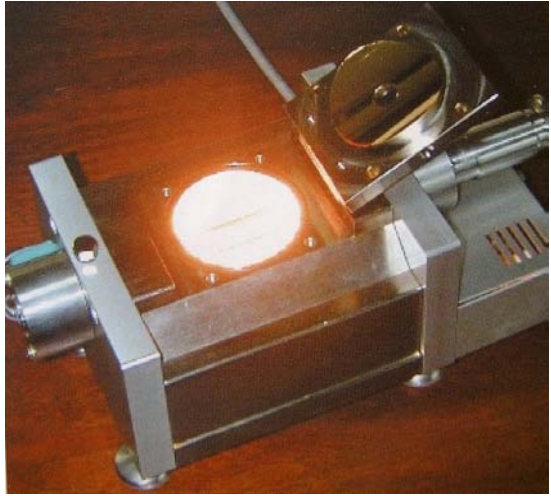

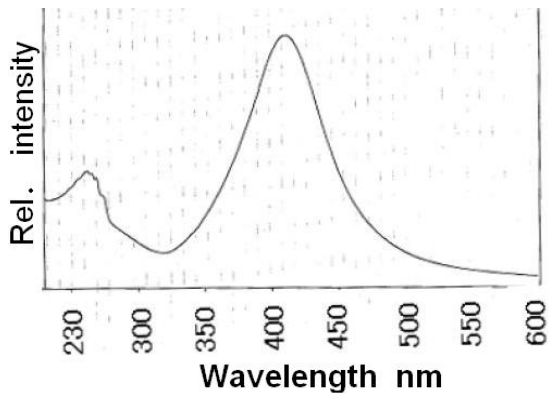

Figure 4. Curing oven (top) and spectral characteristic (bottom) of the electromagnetic radiation used for the curing.

**Electrical tests**

Printable electronics can be used to produce internal wiring, transmission lines, antennas, and theoretically all necessary metals needed in handheld devices. However, the main concern form the electrical point of view is the electrical losses of printed metal traces, especially when application is at high radio frequencies (RF).

The electric losses of transmission lines depend on the conductivity of the material, the trace thickness, and surface roughness of the trace. To determine conductor losses of printed Ag-nano traces deposited by ink jet, we designed and manufactured 50 Ω transmission lines, as shown in Figure 5. Transmission lines were manufactured on ceramic substrate. Two transmission lines were taken under tests. One was manufactured with conventional PCB process by etching (Cu trace created), and another line was realized by Ag-nano ink jet printing. The printed Ag-trace was sintered with profile as 220°C degrees kept for 60 min (by means of air-to-air heating chamber).

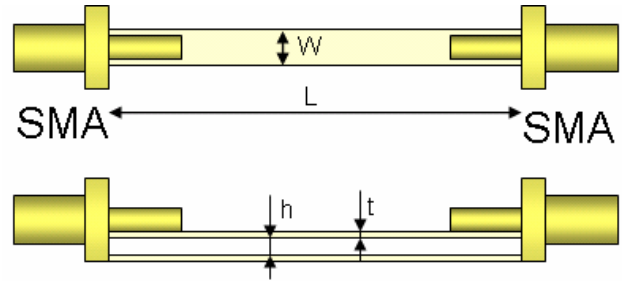

Figure 5. Device under tests.

| Parameter | Value | Unit |
|---|---|---|
| Line width (W) | 3.4 | mm |
| Line length (L) | 60 | mm |
| Height of substrate (h) | 1.62 | mm |
| Thickness of Cu lines (t) | 17 | μm |
| Thickness of Ag lines (t) | 2.0 | μm |
| Dielectric constant | 3.38 | |
| Loss tangent | 0.0021 | |

Table 1. Geometrical and material parameters.

Measurements were done from 50 MHz to 5 GHz with HP8722D Network Analyzer. Figure 6 shows the insertion loss of transmission lines. Solid line represents conventional PCB line and dashed line represents printed Ag-nano ink traces.

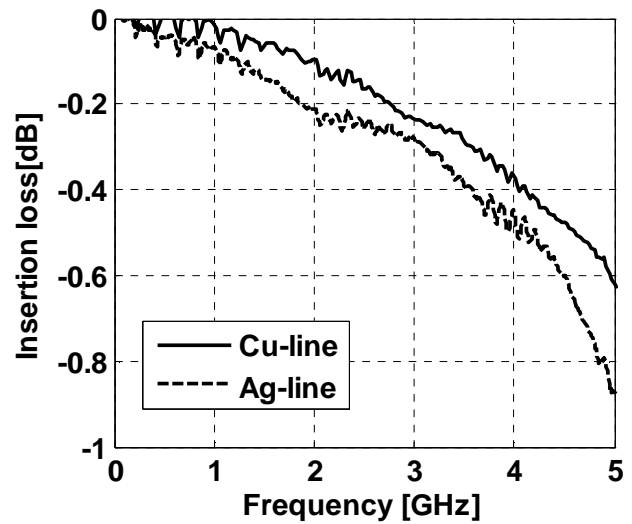

Figure 6. Insertion loss of thick Cu line and thin Ag-line.

From Figure 6 it can be seen that Ag-nano ink traces has greater losses than the Cu lines due to its thinner profile and lower electric conductivity. However, the difference between insertion losses is not so large and can be tolerated in some applications. Therefore, the results indicate that printed Ag-nano traces have relatively good potential for future RF- circuit applications.

## NUMERICAL SIMULATIONS

Thermal and stress analysis has been done by performing thermo-mechanical Finite Element (FE) simulations. A 2D-transient axisymmetric model was made by ABAQUS. The model consisted of an ink jet printed pad on a polymeric layer under which either a polymer layer, Fig 5, or silicon layer could be place. The bottom surface of the model was always connected to a heat sink and a heat transfer coefficient of 10.000 W/m$^2$K was applied. The model was defined so that the temperature of the pad was set to 300$^o$C during 30 s and the initial state of the other parts of the model was set to 20$^o$C. The thermal gradient after 30 s was then calculated for four cases; A – pad under which both of the layers are made of a polymeric material, B – same as case A but heat transfer is also applied around the pad at the top surface, C– pad under which top layer is made of a polymeric material and bottom is layer is made of silicon, and, D – corresponds to case C except that an additional cooling was applied at the top surface around the pad. The calculated thermal gradients below the pad are given in Fig 5 for the different cases. After the temperature distribution was calculated the stresses distribution was in the model. One example of thermal gradient across the assembly is given in Fig 6 for the cases A and B. Also, attributed stress distribution is given at lower part in Fig 6.

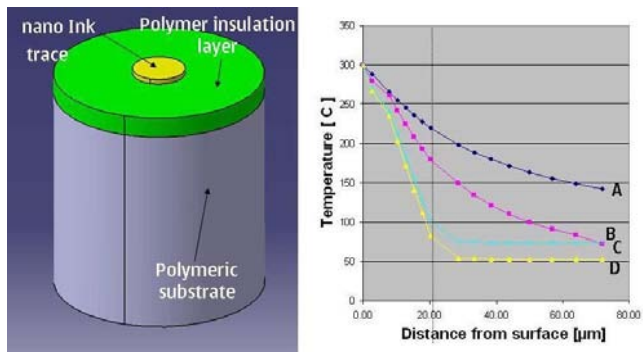

Figure 5. Left: characteristic node (cylindric geometry) consisting of nano-ink trace, an insulation layer and polymeric substrate. Right: Temperature gradient (top-bottom) across the assembly.

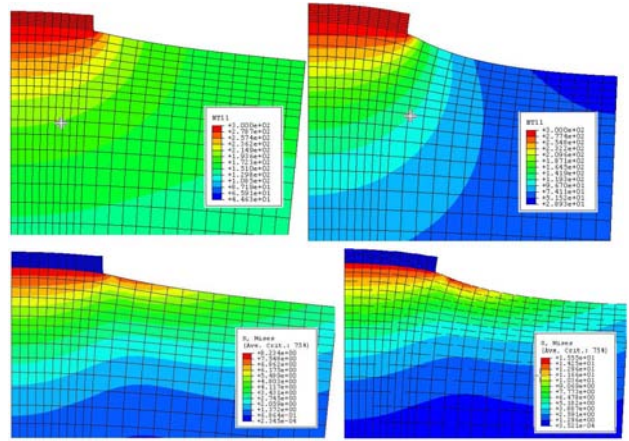

Figure 6. Top: Temperature distribution obtained by ABAQUS simulations for hard curing profile (e.g. T=300$^o$C) in case of additional cooling from the top and top+bottom sides. Bottom: related stress distribution.

It was for example noticed that in case of the "hard" curing profile the substrate will experience relatively high temperatures. For instance at distance of about 20 microns below the assembly surface the temperature is still about 75% of the surface temperature in case A (see Fig 5). Addition cooling paths can slightly minimize the temperature. The draw back of the reduced temperature is that that the thermally induced stresses will increase because of larger thermal gradients. Namely, an additional thermal path introduced from the device bottom (thermal management via bottom heat sink) increased the thermo-mechanical stresses while curing from 8 MPa to 16 MPa (see Fig. 6). Several additional heat propagation conditions have also been evaluated in purpose to examine suitable environment to minimise overall stresses and thermal loadings.

## ANALYSIS AND DISCUSSION

Sintering is one of the most fundamental phenomena in powder metallurgy. Many studies have been made on neck growth during early and later stages of the coalescence and densification at elevated temperatures. The sintering is dependant on many parameters such as particle size, atmospheric conditions (gas type and pressure), temperature and its duration. It is well known that the critical temperature above which densification begins decreases with decrease of particle size [4] (as shown in Fig 7).Usually industrial sintering metallurgies have used approximately ¾ of melting temperatures when compacting the powder metals (mean diameter of about 1 µm). Furthermore, it is well known that at microscopic scale the increase of the atmospheric pressure increase the compacting pressure, decreases the porosity and lowers the melting temperature. However, in the case of Ag-nano-inks (with mean diameter of 5 nm) quantum phenomena dominate the inter-molecular interactions yielding in significant decrease of the sintering temperature (as depicted in Fig 7). This feature might be very useful for coalescence growth and sintering phenomena of nano-inks used for printed electronics.

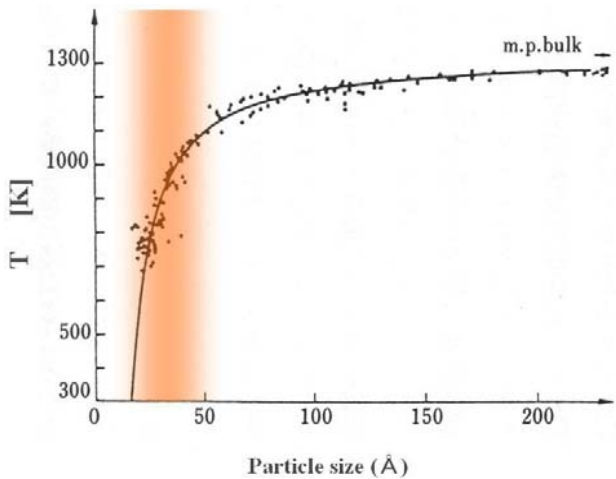

Figure 7. Quantum effects which drastically lower the melting temperature of Ag- nano-particles [4].

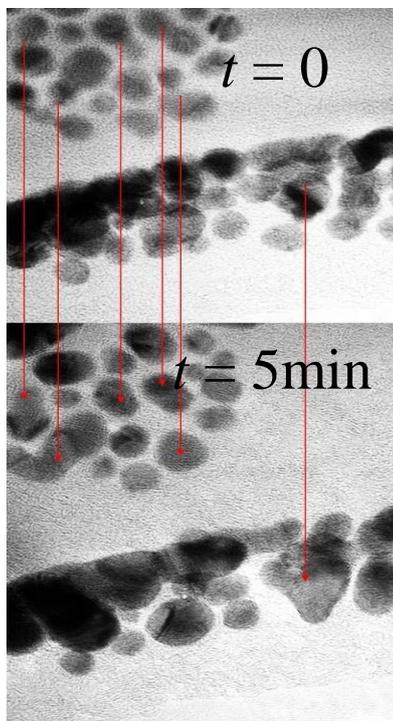

Figure 8. Sintering behavior of the Ag-nano particles vs. time at at elevated temperature.

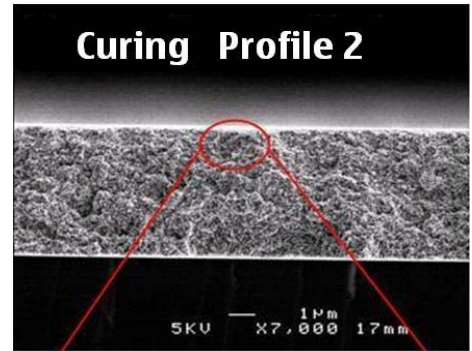

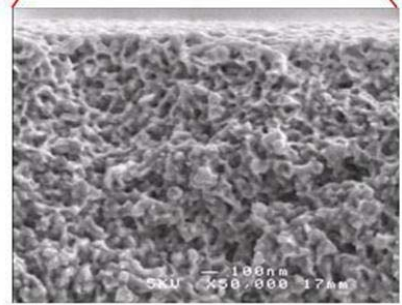

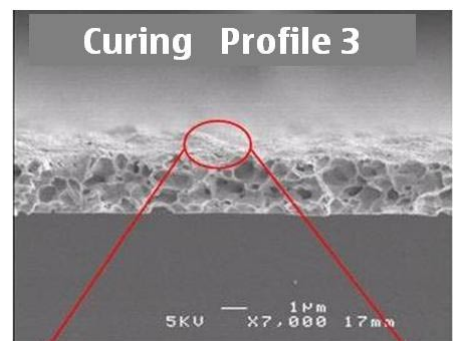

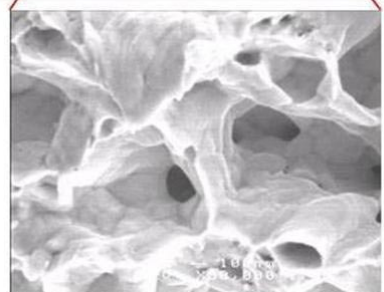

Figure 9. Result of sintering of Ag-nano ink by using different curing profiles. Top: sintering profile 2 consisted of 240°C for 60 min. Bottom: sintering profile 3 consisted of 150°C for 60 min followed by 240°C for 60 min.

The samples were examined over the time of sintering. Micro photographs have been taken sequentially and after exposure to elevated temperature for certain period of time. Fig 8 shows time development of the coalescence of the Ag-nano particles dissolved in the nano-ink.

It is shown that the sintered traces are formed through the coalescence and grain growth of the nano particles.

The quality of the curing was investigated by using structural analysis (TEM microscopy) and 4 point-volumetric resistance measurements. Among several curing profiles the results of the curing with thin samples (<10 µm) by using Profile 2 (permanent temperature of T=240$^o$C for 60 min) showed to be the most optimal. This profile resulted in the most homogeneous and stable cured structures with relatively high electrical conductance of about 3 µΩ-cm). Relative increase of the thickness of the sample (> 10 µm) when curing with the same profile yielded slight increase in specific resistance of about 15%. However the structural behavior seemed to be very similar. The results are shown in Fig 9. In case of using curing profile 3 (pre-heating profile) larger cavities were formed in the cured nano-ink structure due to fluid dynamics phenomena caused larger cavities in the structure of the cured nano-ink. This indicated that the pre-heating temperature was too high and that dynamics of evaporation of the ink solvent made significant structural changes seen as increased porosity. Such porosity weakens the structure and is a risk especially if the final arrangement is made for so called "flexible" electronic assemblies exposed to frequent bending/twisting.

## CONCLUSION AND RECOMMENDATION

Curing of Ag-nano ink based printed traces by using Infra Red technique has been investigated for purpose of optimal curing conditions. The variance of curing conditions were made for a set of printed traces involving maximum curing temperature, time of exposure and introduction of additional cooling paths for the time of curing. It was found that homogeneous curing of the Ag-nano ink can be achieved for thin layer printed lines (<10 µ). In case of thicker layers the influence of out-gassing and sintering fluid dynamics are taking a role which needs to be taken into account. Some suggestions for thicker printed layers include appropriate thermal pre-treatment (preheating at temperatures of about 100$^o$C) and variation of the environmental pressure while curing.

For the thin printed traces the optimal curing profile corresponds to conditions at elevated temperature of about 240$^o$C and for about 60 minutes. In this case the heat treatment does not overstress the whole assembly while it still provides sufficient homogenous sintering and densification of the metallic Ag-nano particles. In other words this sintering profile provides relatively low stresses and result in sufficient electrical conductance of the thin traces. This is an important requirement for functional electronic parts with sufficient length of the routing lines (most of portable electronics applications, e.g functional and peripheral modules).

Finite element simulation results showed that the introduction of external heat path while curing might be a way to minimize the stresses further. It was shown that advanced thermal management needs to be taken from a single side, e.g. from the bottom side of the assembly. Introduction of multidirectional additional heat paths increased the stresses and did not provide "easier" curing condition experienced by the whole assembly.

## ACKNOWLEDGMENT

Authors wish to thank Finish founded Agency for Technology and Innovations (TEKES) for founding and encouraging support for this research.